\PassOptionsToPackage{table,xcdraw}{xcolor}
\documentclass[10pt,conference]{IEEEtran}
\IEEEoverridecommandlockouts

\usepackage{dingbat}
\usepackage{listings}
\usepackage{color}
\usepackage{xcolor}
\usepackage{multirow}
\usepackage{colortbl}
\usepackage{enumitem}
\usepackage{cite}
\usepackage{adjustbox}
\usepackage{amsfonts}
\usepackage{subcaption}
\usepackage[linesnumbered,ruled,vlined]{algorithm2e}
\SetAlgorithmName{Algorithm}{Algorithm}{Algorithm}

\SetAlgoVlined %
\SetAlgoLined %
\SetKwInput{KwInput}{Input}  %
\SetKwInput{KwOutput}{Output}  %
\SetKwProg{Proc}{Procedure}{}{}

\DontPrintSemicolon
\usepackage{graphicx}
\usepackage[most]{tcolorbox}

\usepackage{algpseudocode}
\usepackage{url}
\usepackage{float}
\usepackage[colorlinks=true, linkcolor=blue, citecolor=violet, urlcolor=blue]{hyperref}
\usepackage{caption}
\usepackage{adjustbox}

\algtext*{EndProcedure}%

\definecolor{mypurple}{HTML}{67379A}
\definecolor{myblue}{HTML}{93AAD8}
\definecolor{myred}{HTML}{EB3323}
\definecolor{myyellow}{RGB}{204,153,0}
\definecolor{mygreen}{RGB}{0,102,0}
\definecolor{greenbg}{rgb}{0.9, 1, 0.9}

\lstdefinestyle{mystyle}{
    backgroundcolor=\color{white},   
    basicstyle=\ttfamily\footnotesize,
    breakatwhitespace=false,         
    breaklines=true,                 
    captionpos=b,                    
    keepspaces=true,                 
    numbers=left,                    
    numbersep=5pt,                  
    showspaces=false,                
    showstringspaces=false,
    showtabs=false,                  
    tabsize=2,
    numberstyle=\tiny\color{black},
    framexleftmargin=5mm, frame=single, framerule=0pt,
    xleftmargin=5mm,
    xrightmargin=3mm,
    framesep=5mm, 
    rulecolor=\color{black},
    postbreak=\mbox{\textcolor{red}{$\hookrightarrow$}\space},
}
\usepackage{amsthm}
\theoremstyle{definition}
 
\theoremstyle{plain}

\usepackage{pifont}
\newcommand{\xmark}{\ding{55}} %

\usepackage{tikz}
\usepackage{amsmath}

\newcommand{\circled}[1]{%
  \tikz[baseline=(char.base)]{
    \node[draw, circle, minimum size=0.6em, inner sep=0pt, line width=0.2pt] (char) {\scriptsize \makebox[1.1em]{#1}};
  }%
}


\usepackage{booktabs}

\usepackage{fvextra}

\DefineVerbatimEnvironment{myverbatim}{Verbatim}{
  breaklines=true,
  breakanywhere=true
}

\makeatletter
\newcommand{\removelatexerror}{\let\@latex@error\@gobble}
\makeatother

\makeatletter
\def\tcb@cnt@datalistautorefname{Listing}
\makeatother

\usepackage{color}
\usepackage{soul}
\usepackage{pifont}
\usepackage{lipsum} %
\usepackage{wasysym}

\newtcolorbox[auto counter]{datalist}[2][]{%
  enhanced,
  breakable,                             %
  float,
  floatplacement=t,
  fonttitle=\bfseries\small,
  fontupper=\small,
  colback=gray!10,                       %
  colframe=black,                        %
  fonttitle=\bfseries,                   %
  title={Listing \thetcbcounter. #2},       %
  title after break={Listing \thetcbcounter. #2 (continued)}, %
  rounded corners,                         %
  boxrule=0.5pt,                         %
  top=2pt, bottom=2pt, left=2pt, right=2pt, %
before=\par\vspace{5pt}\noindent,     %
  after=\par,               %
  before upper={\setlength{\parindent}{0pt}}, %
  width=\linewidth,                      %
  #1                                     %
}

\usepackage{nameref} %

\begin{document}

\title{Propagation-Based Vulnerability Impact Assessment for Software Supply Chains}

\author{\IEEEauthorblockN{
Bonan Ruan \hspace{0.3cm}
Zhiwei Lin \hspace{0.3cm}
Jiahao Liu\IEEEauthorrefmark{1}\hspace{0.3cm}
Chuqi Zhang \hspace{0.3cm}
Kaihang Ji \hspace{0.3cm}
Zhenkai Liang}
\IEEEauthorblockA{National University of Singapore}
\{r-bonan, zhiweil, jiahao99, chuqiz, kaihang, liangzk\}@comp.nus.edu.sg
\thanks{\IEEEauthorrefmark{1}Corresponding author}
}

\maketitle

\begin{abstract}
Identifying the impact scope and scale is critical for software supply chain vulnerability assessment.
However, existing studies face substantial limitations.
First, prior studies either work at coarse package-level granularity—producing many false positives—or fail to accomplish whole-ecosystem vulnerability propagation analysis.
Second, although vulnerability assessment indicators like CVSS characterize individual vulnerabilities, no metric exists to specifically quantify the dynamic impact of vulnerability propagation across software supply chains.
To address these limitations and enable accurate and comprehensive vulnerability impact assessment, we propose a novel approach: (i) a hierarchical worklist-based algorithm for whole-ecosystem and call-graph-level vulnerability propagation analysis and (ii) the Vulnerability Propagation Scoring System (VPSS), a dynamic metric to quantify the scope and evolution of vulnerability impacts in software supply chains.
We implement a prototype of our approach in the Java Maven ecosystem and evaluate it on 100 real-world vulnerabilities. 
Experimental results demonstrate that our approach enables effective ecosystem-wide vulnerability propagation analysis, and provides a practical, quantitative measure of vulnerability impact through VPSS.
\end{abstract}

\newcommand{\vpsga}{\textit{GA}\xspace}
\newcommand{\vpsgav}{\textit{GAV}\xspace}
\newcommand{\vpsp}{\textit{P}\xspace}
\newcommand{\vpspv}{\textit{PV}\xspace}

\section{Introduction}
\label{sec:intro}

The proliferation of software vulnerabilities has introduced significant security risks~\cite{zhang2024hitchhiker}.
However, not all vulnerabilities carry the same impact scope.
Specifically, a vulnerability in a client application usually only affects the application itself, while a vulnerability in software supply chains often puts downstream software that depends on the vulnerable upstream libraries directly or transitively at risk as well.
For instance, {\em Heartbleed}~\cite{mitreCVE20140160}, a vulnerability in the OpenSSL library, affects countless services that rely on it.
Another vulnerability, \textit{Log4Shell}, endangers numerous projects depending on the popular Apache Log4j logging framework~\cite{csaLog4ShellVulnerability}.
Hence, it is critical to {\em identify the impact scope and scale of the vulnerability in software supply chains} after a vulnerability is disclosed, which is called \textit{vulnerability propagation analysis}.

Researchers have conducted several studies to investigate this problem for popular programming language ecosystems (\textit{e.g.,} Java~\cite{cadariu2015tracking,plate2015impact,kula2018developers,du2018refining,ponta2018beyond,pashchenko2018vulnerable,hu2019open,wang2020empirical,ponta2020detection,zhang2023mitigating,wu2023understanding,mir2023effect,ma2024vulnet,zhang2024does,shen2025understanding}, JavaScript~\cite{lauinger2018thou,decan2018impact,zimmermann2019small,wang2023plumber,liu2022demystifying}, and Python~\cite{ma2020impact}) by analyzing components and the corresponding dependencies in software supply chains.
While existing works take significant steps toward software supply chain vulnerability analysis, a substantial gap remains in enabling accurate and comprehensive impact assessment.
We observe the following fundamental limitations that render current solutions suboptimal:

\textbf{Limitation 1 (L1):} The lack of accurate and complete vulnerability propagation analysis.
First, many studies only conduct package-level vulnerability propagation analysis based on dependency declarations.
This often leads to false positives, as downstream projects may declare dependencies on vulnerable packages without actually invoking the vulnerable functions (VFs)~\cite{wu2023understanding,shen2025understanding,wang2020empirical}.
Second, although the remaining studies 
have explored call graph (CG)-level analysis, their methods are often limited in scope and incomplete due to a lack of efficient processing techniques for complex whole-ecosystem dependencies.
Specifically, (i) they consider only partial dependency relationships rather than the complex, ecosystem-wide structure;
(ii) they analyze only a subset of project versions or a limited number of downstream projects, instead of covering all relevant versions and dependencies;
and (iii) they focus solely on direct dependencies, ignoring vulnerability propagation through transitive dependencies.
We provide a detailed discussion of these limitations in \autoref{sec:limit-prior-work}.

\textbf{Limitation 2 (L2):}  The lack of metrics for quantifying the impact of vulnerabilities across software supply chains.
The widely adopted vulnerability assessment indicators are only used to characterize the impact of a vulnerability itself.
For instance, although people can perceive the severity of a vulnerability through its CVSS score~\cite{firstCVSSV40FAQ}, it is fundamentally designed to assess and reflect the characteristics of individual vulnerabilities. 
As such, its application does not extend well to measuring vulnerability impacts across software supply chains, which is explicitly acknowledged in the CVSS v4.0 FAQ~\cite{firstCVSSV40FAQ}.

In this work, to address the aforementioned limitations, we propose a novel approach to accurate and whole-ecosystem impact assessment of software supply chain vulnerabilities and a propagation-based indicator for quantifying such impact.

\begin{itemize}[leftmargin=*]
\item 
To address \textbf{L1}, we draw inspiration from data-flow analysis~\cite{khedker2017data,jiang2025fuzzing} and design a worklist-based vulnerability propagation analysis algorithm to efficiently identify affected downstream dependencies given a vulnerability.
The algorithm conducts CG-level analysis that considers complete dependency relations, all potentially affected downstreams, and transitive dependencies across the entire ecosystem.
We integrate hierarchical pruning strategies into the propagation algorithm to reduce the complexity of analysis.

\item 
To address \textbf{L2}, we propose the {\em Vulnerability Propagation Scoring System} (VPSS), a graph-theoretic dynamic indicator specialized for quantifying vulnerability impact in software supply chains and reflecting the temporal evolution of impact.
It is designed to consider both the breadth and depth of vulnerability propagation.
VPSS has a similar score range (0--10) and impact levels (low, medium, high, and critical) as CVSS~\cite{firstCVSSV40FAQ}, making it easy to understand and use.
\end{itemize}

We implement a prototype of our approach for the Java Maven ecosystem~\cite{apacheCentralIndex}, one of the largest software ecosystems in the world, and evaluate it on 100 real-world vulnerabilities investigated by prior work~\cite{wu2023understanding}.
To assess our approach, we ask and answer two evaluation questions: 

\begin{itemize}[leftmargin=*]
\item 
{\em How effective and scalable is our ecosystem-scale vulnerability propagation analysis in identifying affected downstream projects?}
 
{\bf Findings.} 
Our approach successfully and efficiently completes the propagation analysis for all 100 vulnerabilities.
On average, 97.8\% of projects and 99.2\% of project version releases are pruned during the analysis, with the longest and average propagation path lengths 
reduced by at least 34.1\% and 29.4\%, respectively.
Importantly, our approach significantly lowers the cost of CG construction by reducing the number of CGs that need to be built.

\item 
{\em What insights can be drawn from the VPSS scores?}

{\bf Findings.}
The computed VPSS scores generally decline over time after disclosure, driven by patch adoption and ecosystem expansion.
Interestingly, some CVEs (\textit{e.g.,} CVE-2016-3086~\cite{nistCVE20163086}) show temporary score increases due to delayed dependency updates. Across the entire dataset, VPSS scores remain relatively low and gradually stabilize.
This aligns with expected ecosystem dynamics where vulnerability propagation attenuates over time.

\end{itemize}

To the best of our knowledge, this is the first work achieving CG-level and whole-ecosystem vulnerability impact assessment for software supply chains.
In summary, this paper makes the following contributions:

\begin{itemize}[leftmargin=*]
\item We design a hierarchical worklist-based vulnerability propagation analysis algorithm to accurately and efficiently identify affected downstream dependencies across a whole software ecosystem.
\item We propose Vulnerability Propagation Scoring System (VPSS), the first time-aware indicator for quantifying vulnerability impact in software supply chains.
\item We implement a prototype of our approach for the Java Maven ecosystem and evaluate it on real-world vulnerabilities. The code and experimental dataset are available at \url{https://github.com/brant-ruan/vpss}.
\end{itemize}

\section{Background and Motivations}

\subsection{Background}\label{sec:background}

\subsubsection{Terminology}

\begin{table}[t]
  \centering
  \caption{Term Unification Across Ecosystems}
  \label{tab:terminology}
\resizebox{0.8\linewidth}{!}{%
  \begin{tabular}{l|l|l}
    \toprule
    \textbf{Ecosystem} & \textbf{Project (\vpsp)} & \textbf{Project-Version (\vpspv)} \\
    \midrule
    Maven (Java) & GroupId:ArtifactId & GroupId:ArtifactId:Version \\
    npm (JavaScript) & package name & package@version \\
    PyPI (Python) & distribution name & name==version \\
    \bottomrule
  \end{tabular}
}
\end{table}

Different software ecosystems use diverse naming conventions to refer to software units and their versions, such as \textit{packages}, \textit{modules}, or \textit{distributions}.
This inconsistency may lead to inaccurate descriptions and hinder understanding.
To provide a unified abstraction across ecosystems, we use the terms \textit{Project} ({\vpsp}) and \textit{Project-Version} ({\vpspv}) to represent software units and their specific releases, respectively. 
\autoref{tab:terminology} shows how these terms correspond to identifiers in representative ecosystems.

\subsubsection{Vulnerability Propagation Analysis}\label{sec:rel-work}

Given a vulnerability, identifying the scope and scale of its impact in software supply chains is called \textit{vulnerability propagation analysis}, which takes vulnerability intelligence and inter-project dependencies as input to reason out the vulnerability's impact on downstream projects.
Vulnerability propagation analysis can be conducted at different granularity levels.
One option is the {\vpspv}-level analysis, which considers a downstream {\vpspv} as affected by the vulnerability if it declares a dependency on the upstream vulnerable {\vpspv}s.
The other one is the call graph (CG)-level analysis, which regards a downstream {\vpspv} as affected only when it directly or transitively calls vulnerable functions (VFs) of the upstream vulnerable {\vpspv}s.
A prerequisite for CG-level analysis is to identify the VFs, where the vulnerable logic exists.
Currently, the most widely adopted VF identification method is the patch-based approach~\cite{dai2020bscout,wu2023understanding,ponta2018beyond,jiang2020pdiff,liu2020large}, which identifies functions deleted or modified in patches as VFs, a widely adopted strategy due to its logical rationale and alignment with standardized patch information.

\subsection{Limitations of Existing Solutions}

\begin{table*}[t]
  \centering
  \caption{Summary of related works in comparison with this work. In `LAN', `JA' stands for Java, `JS' stands for JavaScript, and `PY' stands for Python. In `Granularity', `PV' means the work analyzes the propagation at {\vpspv} level, and `CG' means CG-level analysis. In `VF Identification', `Manual' means the work identifies VFs manually, `Patch' means the work uses a patch-based method to identify VFs, and `Patch (Optimized)' means the work uses an optimized patch-based method.}
  \resizebox{0.80\textwidth}{!}{%
  \begin{tabular}{c|c|l|c|c|c|c|c|c}
    \toprule
    Year & LAN & Research & Direction & Dep Scope & Coverage & Transitivity & Granularity & VF Identification \\
    \midrule
    2015 & JA & Cadariu \textit{et al.}~\cite{cadariu2015tracking}
        & \textcolor{myyellow}{Forward}   & \textcolor{myyellow}{Partial}   & \textcolor{myyellow}{Partial}   & \textcolor{myyellow}{Direct}     & \textcolor{myyellow}{PV}         & \textcolor{myyellow}{\xmark}              \\
    2015 & JA & Ponta \textit{et al.}~\cite{plate2015impact}
        & \textcolor{myyellow}{Forward}   & \textcolor{myyellow}{Partial}   & \textcolor{myyellow}{Partial}   & \textcolor{myyellow}{Direct}     & \textcolor{myyellow}{PV}         & Patch          \\
    2017 & JS & Lauinger \textit{et al.}~\cite{lauinger2018thou}
        & \textcolor{myyellow}{Forward}   & \textcolor{myyellow}{Partial}   & \textcolor{myyellow}{Partial}   & \textcolor{mygreen}{Transitive} & \textcolor{myyellow}{PV}         & \textcolor{myyellow}{\xmark}              \\
    2018 & JS & Decan \textit{et al.}~\cite{decan2018impact}
        & \textcolor{myyellow}{Forward}   & \textcolor{myyellow}{Partial}   & \textcolor{myyellow}{Partial}   & \textcolor{myyellow}{Direct}     & \textcolor{myyellow}{PV}         & \textcolor{myyellow}{\xmark}              \\
    2018 & JA & Kula \textit{et al.}~\cite{kula2018developers}
        & \textcolor{myyellow}{Forward}   & \textcolor{myyellow}{Partial}   & \textcolor{myyellow}{Partial}   & \textcolor{myyellow}{Direct}     & \textcolor{myyellow}{PV}         & \textcolor{myyellow}{\xmark}              \\
    2018 & JA & Du \textit{et al.}~\cite{du2018refining}
        & \textcolor{myyellow}{Forward}   & \textcolor{myyellow}{Partial}   & \textcolor{myyellow}{Partial}   & \textcolor{myyellow}{Direct}     & \textcolor{myyellow}{PV}         & \textcolor{myyellow}{\xmark}              \\
    2018 & JA & Ponta \textit{et al.}~\cite{ponta2018beyond}
        & \textcolor{myyellow}{Forward}   & \textcolor{myyellow}{Partial}   & \textcolor{myyellow}{Partial}   & \textcolor{myyellow}{Direct}     & \textcolor{mygreen}{CG}   & Patch          \\
    2018 & JA & Pashchenko \textit{et al.}~\cite{pashchenko2018vulnerable}
        & \textcolor{myyellow}{Forward}   & \textcolor{myyellow}{Partial}   & \textcolor{myyellow}{Partial}   & \textcolor{myyellow}{Direct}     & \textcolor{myyellow}{PV}         & Patch          \\
    2019 & JA & Hu \textit{et al.}~\cite{hu2019open}
        & \textcolor{myyellow}{Forward}   & \textcolor{myyellow}{Partial}   & \textcolor{mygreen}{Complete}   & \textcolor{mygreen}{Transitive}  & \textcolor{myyellow}{PV}         & \textcolor{myyellow}{\xmark}              \\
    2019 & JS & Zimmermann \textit{et al.}~\cite{zimmermann2019small}
        & \textcolor{mygreen}{Backward}  & \textcolor{mygreen}{Complete}   & \textcolor{mygreen}{Complete}   & \textcolor{mygreen}{Transitive}  & \textcolor{myyellow}{PV}         & \textcolor{myyellow}{\xmark}              \\
    2020 & JA & Wang \textit{et al.}~\cite{wang2020empirical}
        & \textcolor{myyellow}{Forward}   & \textcolor{myyellow}{Partial}   & \textcolor{myyellow}{Partial}   & \textcolor{myyellow}{Direct}     & \textcolor{mygreen}{CG}   & Patch          \\
    2020 & JA & Ponta \textit{et al.}~\cite{pashchenko2020vuln4real,ponta2020detection}
        & \textcolor{myyellow}{Forward}   & \textcolor{myyellow}{Partial}   & \textcolor{myyellow}{Partial}   & \textcolor{myyellow}{Direct}     & \textcolor{mygreen}{CG}   & Patch          \\
    2020 & PY & Ma \textit{et al.}~\cite{ma2020impact}
        & \textcolor{mygreen}{Backward}  & \textcolor{myyellow}{Partial}   & \textcolor{myyellow}{Partial}   & \textcolor{mygreen}{Transitive}  & \textcolor{mygreen}{CG}   & \textcolor{myyellow}{Manual}              \\
    2022 & JS & Liu \textit{et al.}~\cite{liu2022demystifying}
        & \textcolor{mygreen}{Backward}  & \textcolor{mygreen}{Complete}   & \textcolor{mygreen}{Complete}   & \textcolor{mygreen}{Transitive}  & \textcolor{myyellow}{PV}         & \textcolor{myyellow}{\xmark}              \\
    2023 & JS & Wang \textit{et al.}~\cite{wang2023plumber}
        & \textcolor{mygreen}{Backward}  & \textcolor{mygreen}{Complete}   & \textcolor{mygreen}{Complete}   & \textcolor{mygreen}{Transitive}  & \textcolor{myyellow}{PV}         & \textcolor{myyellow}{\xmark}              \\
    2023 & JA & Zhang \textit{et al.}~\cite{zhang2023mitigating}
        & \textcolor{mygreen}{Backward}  & \textcolor{mygreen}{Complete}   & \textcolor{mygreen}{Complete}   & \textcolor{mygreen}{Transitive}  & \textcolor{myyellow}{PV}         & \textcolor{myyellow}{\xmark}              \\
    2023 & JA & Wu \textit{et al.}~\cite{wu2023understanding}
        & \textcolor{myyellow}{Forward}   & \textcolor{mygreen}{Complete}   & \textcolor{myyellow}{Partial}   & \textcolor{myyellow}{Direct}     & \textcolor{mygreen}{CG}   & Patch          \\
    2023 & JA & Mir \textit{et al.}~\cite{mir2023effect}
        & \textcolor{myyellow}{Forward}   & \textcolor{myyellow}{Partial}   & \textcolor{mygreen}{Complete}   & \textcolor{mygreen}{Transitive}  & \textcolor{mygreen}{CG}   & Patch          \\
    2024 & JA & Ma \textit{et al.}~\cite{ma2024vulnet}
        & \textcolor{myyellow}{Forward}   & \textcolor{myyellow}{Partial}   & \textcolor{mygreen}{Complete}   & \textcolor{mygreen}{Transitive}  & \textcolor{myyellow}{PV}         & \textcolor{myyellow}{\xmark}              \\
    2024 & JA & Zhang \textit{et al.}~\cite{zhang2024does}
        & \textcolor{mygreen}{Backward}  & \textcolor{myyellow}{Partial}   & \textcolor{mygreen}{Complete}   & \textcolor{myyellow}{Direct}     & \textcolor{mygreen}{CG}  & Patch (Optimized) \\
    2025 & JA & Shen \textit{et al.}~\cite{shen2025understanding}
        & \textcolor{mygreen}{Backward}  & \textcolor{myyellow}{Partial}   & \textcolor{myyellow}{Partial}   & \textcolor{mygreen}{Transitive}  & \textcolor{mygreen}{CG}   & Patch          \\
    \midrule
         &    & \textbf{This Work}
        & \textcolor{mygreen}{Backward}  & \textcolor{mygreen}{Complete} & \textcolor{mygreen}{Complete}   & \textcolor{mygreen}{Transitive}  & \textcolor{mygreen}{CG}   & Patch (Optimized) \\
    \bottomrule
  \end{tabular}%
  }
  \label{tab:rel-work}
\end{table*}

\subsubsection{Vulnerability Propagation Analysis}\label{sec:limit-prior-work}

To better profile existing vulnerability propagation research, we conduct a comprehensive literature review.
Prior works are mainly empirical studies on Java~\cite{cadariu2015tracking,plate2015impact,kula2018developers,du2018refining,ponta2018beyond,pashchenko2018vulnerable,hu2019open,wang2020empirical,ponta2020detection,zhang2023mitigating,wu2023understanding,mir2023effect,ma2024vulnet,zhang2024does,shen2025understanding}, JavaScript~\cite{lauinger2018thou,decan2018impact,zimmermann2019small,wang2023plumber,liu2022demystifying}, and Python~\cite{ma2020impact} ecosystems, as shown in
\autoref{tab:rel-work}.
To clearly display and compare these works, we profile them from six aspects: 

\textit{Direction} indicates whether the work conducts a forward analysis to answer \textit{`which vulnerabilities in upstream dependencies affect a downstream project'}, or a backward analysis to answer \textit{`which downstream projects (as the call sites) are affected by an upstream vulnerability---by using a chain of function calls to reach it'}.
Intuitively, backward analysis is more suitable for vulnerability impact analysis, as it starts from the vulnerable site and propagates to the downstream projects.

\textit{Dep Scope} clarifies whether the work examines the partial or complete dependency relations for a target ecosystem.
For example, if a work only selects a subset from a software ecosystem with their dependencies for analysis, it has partial dependency scope.
Instead, if the work conducts the propagation analysis with considering the whole ecosystem and dependency relations, the dependency scope is complete.

\textit{Coverage} shows whether the work analyzes partial or complete projects.
For example, if a work only selects one version to represent the target project, then it has partial coverage.
Conversely, if the work considers all released versions for the propagation, it has complete coverage.

\textit{Transitivity} indicates whether the work analyzes only the direct dependency relations or the transitive dependencies.
Direct analysis only takes projects that directly depend on the vulnerable project into consideration, \textit{i.e.,} one-hop dependency;
transitive analysis considers multi-hop dependency towards the root vulnerable project.

\textit{Granularity} shows the granularity at which a work conducts the propagation analysis.
CG-level analysis inspects whether downstream {\vpspv}s directly or transitively call upstream VFs, which is much more accurate than {\vpspv}-level analysis that only considers {\vpspv} dependency relations~\cite{wu2023understanding}.

{\em VF identification} indicates whether the work identifies VFs and how it identifies them.
Manual identification cannot be scaled to large ecosystems.
Although the patch-based method has been widely used in existing work, it has two limitations.
First, patches are not always publicly available~\cite{xu2022tracking}, though several methods have been proposed to address this~\cite{zhou2021spi,tan2021locating,xu2022tracking,wang2022vcmatch,dunlap2024vfcfinder,yu2024llm}, which are beyond the scope of this paper.
Second, patches sometimes contain changes unrelated to the vulnerability.

\subsubsection{Vulnerability Assessment}

For a software supply chain vulnerability, it is equally important to assess its own characteristics and its impact on downstream dependents.
The Common Vulnerability Scoring System (CVSS) provides a way to capture the principal characteristics of a vulnerability and produce a numerical score (0--10) reflecting its severity~\cite{cvss}.
However, it is fundamentally designed to assess and reflect the characteristics and severity of individual vulnerabilities, and does not extend well to measuring vulnerability impacts across software supply chains. 
This limitation is explicitly acknowledged in the CVSS v4.0 FAQ~\cite{firstCVSSV40FAQ}, where it is clarified that \textit{there is no prescribed way to use CVSS Base and Environmental metrics to score a vulnerability along a long supply chain}.
Furthermore, prior research mainly focuses on automating existing assessments~\cite{han2017learning,elbaz2020fighting,le2021deepcva,le2022use,aghaei2023automated,pan2024towards,pan2023fine,fu2023vulexplainer,wen2024livable,luo2024predicting,ji2024applying} or proposing new metrics~\cite{bozorgi2010beyond,sabottke2015vulnerability,tavabi2018darkembed,chen2019using,jacobs2020improving,jacobs2021exploit,suciu2022expected,jacobs2023enhancing} to profile a vulnerability's own characteristics.
How to assess the impact of a vulnerability in software supply chains is still an open problem.

Overall, for vulnerability propagation analysis, an accurate and comprehensive solution should be a backward, transitive, CG-level analysis that considers complete dependency scope and project coverage and is capable of identifying VFs.
Nevertheless, according to our comprehensive investigation, there still remains a gap between existing works and this goal.
For vulnerability assessment, the community needs a new metric reflecting vulnerability impact in software supply chains.

\begin{figure}[t]
  \centering
  \includegraphics[width=\linewidth]{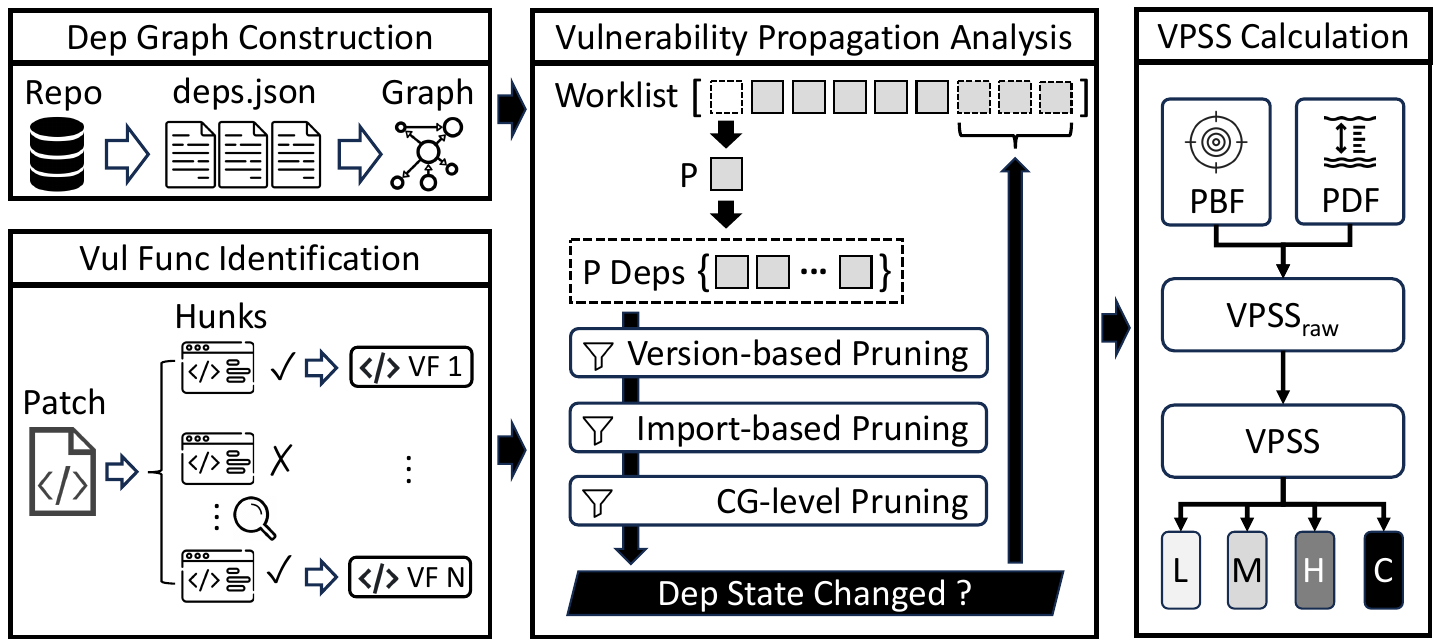}
  \caption{Approach Overview}
  \label{fig:overview}
\end{figure}

\section{Approach}\label{sec:approach}

\subsection{Overview}\label{sec:appro_overview}

\autoref{fig:overview} illustrates the overview of our approach, which is designed to be ecosystem-agnostic and can be adapted to various programming languages by incorporating ecosystem-specific metadata formats and program analysis tools.
This approach is composed of four steps:

{\bf Dependency Graph Construction (\autoref{sec:dep_graph}).} 
To carry out vulnerability propagation analysis, we first need to identify all the downstream projects depending on the upstream project where the given vulnerability is located.
We construct a \vpsp-level dependency graph for this purpose by analyzing all dependency declaration files from the target software ecosystem.

{\bf Vulnerable Function Identification (\autoref{sec:vf_identify}).}
For CG-level analysis, VFs serve as the starting points.
This step first generates a list of VF candidates using the patch-based method, and then takes a large language model (LLM)-assisted strategy to filter out vulnerability-irrelevant candidates.

\textbf{Vulnerability Propagation Analysis (\autoref{sec:vuln_propagate}).}
This step is to effectively identify all the downstream {\vpspv}s in the whole ecosystem that directly or transitively call the VFs in the root vulnerable upstream {\vpspv}s.
We design a hierarchical worklist-based propagation algorithm to achieve this goal.

\textbf{VPSS Calculation (\autoref{sec:vpss}).}
This step calculates the VPSS score based on the results of the propagation analysis, which considers both the breadth and depth of the vulnerability impact scope in software supply chains.

\subsection{Dependency Graph Construction}\label{sec:dep_graph}

Given a vulnerability and the {\vpspv}s affected by it, one of the preliminaries is to figure out which {\vpspv}s depend on these vulnerable {\vpspv}s.
The graph structure can effectively organize {\vpspv}s and their dependency information for querying.
We call this directed graph a dependency graph.
Based on the level of granularity, the dependency graph can be constructed in two distinct ways: the {\vpspv}-level and the {\vpsp}-level design.
In the {\vpspv}-level dependency graph, nodes represent {\vpspv}s and edges denote direct dependencies between them. In contrast, the {\vpsp}-level graph abstracts nodes as {\vpsp}s, with edges summarizing relations derived from the underlying {\vpspv}-level dependencies.

Existing graph-based studies all construct {\vpspv}-level dependency graphs.
However, based on two observations, this option is not efficient for ecosystem-scale vulnerability propagation analysis.
First, modern software ecosystems have become extremely large.
For example, the Java Maven ecosystem has more than 15 million {\vpspv}s~\cite{mvnrepositoryMavenRepository}, and the {\vpspv}-level dependency graph could have tens of millions of nodes and even more edges.
This huge scale makes the {\vpspv}-level dependency graph too large to be queried efficiently, especially for transitive dependencies.
Second, the number of {\vpspv}s that depend on the upstream vulnerable {\vpspv}s is usually a minority in a target ecosystem. 
Even among the {\vpspv}s that do have dependencies, the ones that are actually affected are not many~\cite{wu2023understanding}, which means that most of the nodes and edges in the {\vpspv}-level dependency graph are irrelevant to the vulnerability propagation analysis.
Therefore, it is not necessary and inefficient to construct a {\vpspv}-level dependency graph for a whole ecosystem.

To address these issues, we propose to construct a {\vpsp}-level dependency graph.
The number of {\vpsp}s in an ecosystem is usually much smaller than that of {\vpspv}s, meaning that the {\vpsp}-level dependency graph has a much smaller scale than the {\vpspv}-level dependency graph.
The {\vpsp}-level graph can not only reduce computational overhead, but also serve as an efficient pre-filter, allowing queries to quickly narrow down the search space.
With the initial traversal confined to a smaller, less complex {\vpsp}-level graph, the overall analysis becomes more scalable.

We follow a four-step procedure to build the {\vpsp}-level dependency graph.
First, we download the ecosystem index and extract all the {\vpspv} identifiers from the index into a list.
Second, we obtain the dependency declaration files for {\vpspv}s in this list from the official repository.
Third, we parse the dependency declaration files to extract the dependencies of each {\vpspv} and save them into \verb|deps.json| files.
Fourth, we construct the {\vpsp}-level dependency graph by analyzing the \verb|deps.json| files of all the {\vpspv}s.
Specifically, for each {\vpsp}, we aggregate all the recorded dependent {\vpspv}s from the \verb|deps.json| files of the {\vpspv}s that belong to it.
It is important to note that we only build the dependency graph once and then continue to update it incrementally following the official repository, rather than building the entire dependency graph each time a new vulnerability is analyzed, which greatly reduces the time cost and improves the  efficiency.
During the propagation analysis (\autoref{sec:vuln_propagate}) for a new vulnerability, a corresponding subgraph is queried, and we conduct a targeted {\vpspv}-level inspection by querying the \verb|deps.json| files when necessary, at which point the scale of the subject has been significantly reduced.

\begin{figure}[t]
  \centering
  \includegraphics[width=0.8\linewidth]{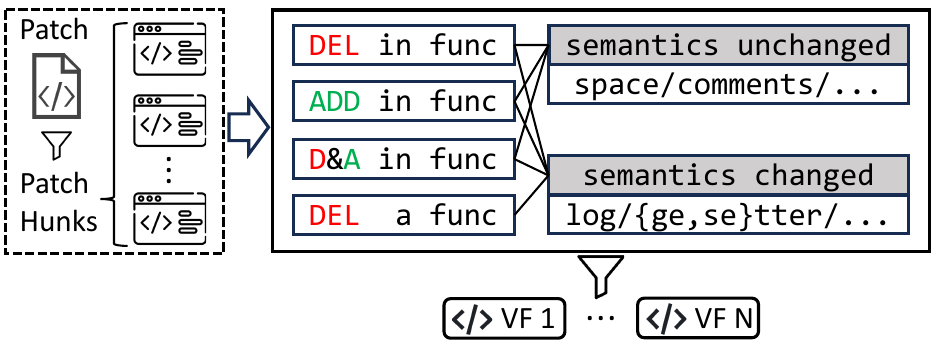}
  \caption{Vulnerable Function Identification}
  \label{fig:vf_identify}
\end{figure}

\subsection{Vulnerable Function Identification}\label{sec:vf_identify}

As presented in \autoref{fig:vf_identify}, the VF identification step consists of two substeps: (1) patch-based VF candidate generation and (2) LLM-assisted VF filtering.

First, we parse the patch of the target vulnerability into individual hunks and only extract the function-modifying hunks.
There are five types of function-modifying hunks: function \textit{addition \& deletion}, and internal \textit{deletion \& addition \& modification (including deletion and addition)}.
We filter out the function addition hunks as prior work~\cite{wu2023understanding} does because they are not the root cause of the vulnerability.

Second, we leverage LLMs for VF filtering.
This method has three advantages:
(1) LLMs possess broad domain knowledge across programming languages, enabling us to develop a language-agnostic and generalizable filtering approach.
(2) LLMs are capable of recognizing non-standard syntax and syntactic sugar that are difficult to enumerate manually.
(3) As LLMs continuously evolve and incorporate newly observed patterns from code corpora, their filtering capabilities remain up-to-date and adaptable, whereas manually maintained rules are often incomplete and costly to update.

Specifically, for the remaining VF hunk candidates, we design an in-context learning (ICL)~\cite{dong2022survey} strategy and drive LLMs to follow two filtering principles: semantics-equivalent modification and semantics-changing modification.
If the semantics of a function remain equivalent after being modified by a hunk, the hunk is considered irrelevant to the vulnerability and should be filtered out, because it does not affect the existence of the vulnerability.
For example, if a hunk only changes variable names, adds or deletes whitespaces, it is likely to be vulnerability-irrelevant.
Even if a hunk changes the semantics of a function, it still can be irrelevant to the vulnerability.
For example, if a hunk only adds or deletes logging or debugging code, it is likely to be irrelevant.
To reduce incorrect filtering caused by LLMs, we also ask LLMs to provide reasons for decisions to conduct manual verification.

\subsection{Vulnerability Propagation Analysis}\label{sec:vuln_propagate}

This section describes the entire procedure of vulnerability propagation analysis.
{\autoref{sec:vpa_overall}} presents its overview that beginning with the root upstream {\vpsp}, each pass of the analysis handles the direct dependencies between the upstream {\vpsp} and its downstream {\vpsp}s.
{\autoref{sec:vpa_prune}} explains how the analysis prunes the downstream {\vpspv}s with hierarchical methods to minimize the number of dependencies.
Furthermore, {\autoref{sec:vpa_workflow}} shows the detailed algorithm workflow.
Last, {\autoref{sec:vpa_example}} illustrates the entire procedure with a real-world example.

\subsubsection{Overall Procedure}\label{sec:vpa_overall}
\begin{figure}[t]
  \centering
  \includegraphics[width=0.6\linewidth]{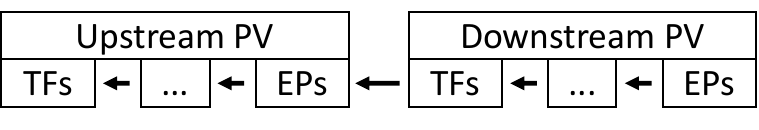}
  \caption{Call Path Illustration}
  \label{fig:up-down-call}
\end{figure}

The vulnerability propagation analysis is to start from a root {\vpsp} (comprising a series of vulnerable {\vpspv}s) with a vulnerability and identify all downstream {\vpsp}s (with the corresponding {\vpspv}s) affected by this vulnerability at CG level along the dependency graph.
Specifically, given an upstream {\vpsp}, we first query the dependency graph to get its direct downstream dependent {\vpsp}s, verify the validity of dependency between each pair of upstream and downstream {\vpsp}s by inspecting whether the downstream {\vpspv}s transitively call the VFs in the root {\vpspv}s, and then recursively propagate the analysis only for the truly affected downstream {\vpsp}s.
\autoref{fig:up-down-call} illustrates this inter-{\vpspv} analysis:
For each {\vpspv} of an upstream {\vpsp} in affected target versions (TVs), we first need to identify the entry-point functions (EPs) that can reach the target functions (TFs) in the same upstream {\vpspv} and be called from outside.
For {\vpspv}s of the root {\vpsp}, VFs are the TFs.
Then, we need to identify which downstream {\vpspv}s call EPs in the upstream {\vpspv}. 
If any, we need to record the functions that call upstream EPs in the downstream {\vpspv}s, which serve as the TFs in future rounds.

\begin{figure}[t]
  \centering
  \begin{subfigure}[t]{0.12\linewidth}
    \centering
    \includegraphics[width=\linewidth]{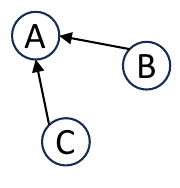}
    \caption{}
    \label{fig:dep-1}
  \end{subfigure}
  \hspace{2em}
  \begin{subfigure}[t]{0.12\linewidth}
    \centering
    \includegraphics[width=\linewidth]{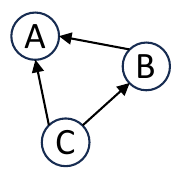}
    \caption{}
    \label{fig:dep-2}
  \end{subfigure}
  \hspace{2em}
  \begin{subfigure}[t]{0.12\linewidth}
    \centering
    \includegraphics[width=\linewidth]{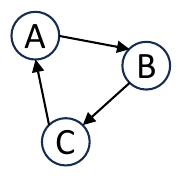}
    \caption{}
    \label{fig:dep-3}
  \end{subfigure}
  \caption{Dependency Scenarios}
  \label{fig:combined-deps}
\end{figure}

In terms of vulnerability propagation, we need to consider three possible dependency scenarios.
In \autoref{fig:dep-1}, downstream B and C have individual dependencies on A.
Therefore, we can identify affected downstream {\vpspv}s for A, B, and C sequentially (\verb|{A,B,C}|).
In \autoref{fig:dep-2}, downstream B and C share a common dependency on A, while C also has a dependency on B.
In this scenario, analysis order \verb|{A,B,C}| and \verb|{A,C,B}| could potentially lead to different results.
The result of \verb|{A,C,B}| is potentially incomplete, because only EPs of upstream A exist when C is being processed, and the algorithm may miss the EPs of upstream B that are called by downstream C.
For \verb|{A,C,B}|, there should be a mechanism to ensure that C is analyzed again after its upstream EPs are updated (\textit{i.e.,} B involves new EPs).
In \autoref{fig:dep-3}, the dependencies form a cycle, which could lead to infinite analysis loops if not handled properly.
Although software dependencies are expected to form a directed acyclic graph~\cite{wikipediaAcyclicDependencies}, cyclic dependency relationships still exist in real-world software ecosystems.
For example, \textit{dom4j:dom4j:1.5.2}~\cite{dom4j} and \textit{jaxen:jaxen:1.1-beta-4}~\cite{jaxen} have mutual dependencies on each other.

To effectively handle the aforesaid scenarios, we adopt the worklist algorithm, a well-established method in data-flow analysis frameworks, to systematically perform vulnerability propagation analysis over the dependency graph. 
Specifically, we maintain a worklist of {\vpsp}s whose states (\textit{i.e.,} TVs, versioned reachable EPs, and affected downstream {\vpspv}s) may still be updated. 
Initially, only the root {\vpsp} is added to the worklist. In each pass, we dequeue an item from the worklist, perform inter-{\vpspv} analysis to update its downstream items, and enqueue the affected downstream {\vpsp}s if their associated versioned reachable upstream EPs have been updated.
This approach ensures that each {\vpsp} is revisited only when necessary, effectively handling shared dependencies and cyclic structures while avoiding redundant analysis.
The propagation continues until a fixed point is reached—when no new updates occur across the graph, ensuring both soundness and efficiency.

\subsubsection{Pruning Mechanism}\label{sec:vpa_prune}
\textbf{Hierarchical Pruning.} We notice that there could be a large number of {\vpspv} dependencies involved in the analysis, and the time and space costs associated with constructing CGs for large {\vpspv}s are non-negligible.
Consequently, it is computationally inefficient to analyze all dependencies directly at the CG level. 
To avoid lots of unnecessary fine-grained analysis, we employ a hierarchical pruning strategy, which first applies coarse-grained pruning methods to efficiently exclude false positive downstream dependent {\vpspv}s (as well as their corresponding {\vpsp}s, if all associated {\vpspv}s are pruned out), and then performs the fine-grained CG-level analysis only on the remaining downstream candidates.
As shown in the middle part of \autoref{fig:overview}, this hierarchical pruning mechanism comprises three levels of pruning:
\textit{Version-based pruning} excludes downstream {\vpspv}s that do not declare a dependency on the specific upstream TVs.
\textit{Import-based pruning} further eliminates downstream {\vpspv}s that, despite declaring a dependency, do not actually import or include upstream contents. 
\textit{CG-level pruning} finally removes downstream {\vpspv}s that do not invoke any of the upstream EPs at the CG level.

\textbf{Handling Fat Packages.}
An issue in the pruning mechanism arises from the presence of fat {\vpspv}s—release packages that bundle not only a project's own code but also its dependencies.
Such packaging practices, common in ecosystems like Java where fat JARs are widely used, can interfere with precise analysis by conflating intrinsic and extrinsic program elements.
For efficient analysis, only the \textit{intrinsic scope} (\textit{i.e.,} the program components that are native to the project itself) should be considered. 
To address this issue, we propose a general method for identifying the intrinsic scope of a given {\vpspv}, even when fat packaging practices differ across ecosystems.
Specifically, for a target {\vpspv}, we first obtain the set of files in its release package, denoted as \textit{Up}.
We then collect the release files of all its declared dependencies as set \textit{Down}.
By subtracting \textit{Down} from \textit{Up}, we obtain the intrinsic scope of the {\vpspv}, which serves as the foundation for import-based and CG-level pruning.

\begin{figure}[t]
  \centering
  \includegraphics[width=0.77\linewidth]{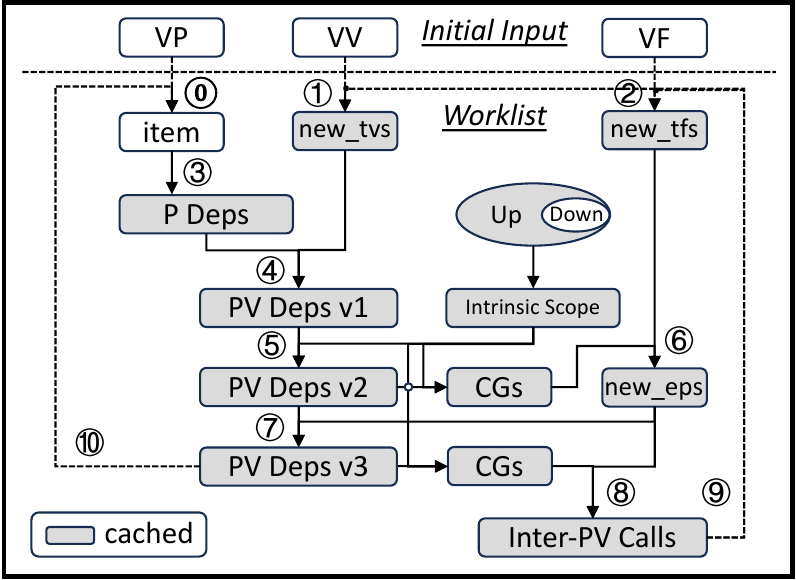}
  \caption{Vulnerability Propagation Analysis}
  \label{fig:vpa-detailed}
\end{figure}

\subsubsection{Algorithm Workflow}\label{sec:vpa_workflow}
Taking all the above into consideration, we design a hierarchical worklist-based algorithm to perform vulnerability propagation analysis, presented in \autoref{fig:vpa-detailed}, with the corresponding steps annotated using circled numbers that match those in the pseudocode provided in \autoref{alg:worklist} for clarity.
Given a vulnerability, the algorithm takes three inputs: the root {\vpsp} of the vulnerability ($\mathcal{VP}$), the vulnerable versions of the root \vpsp ($\mathcal{VV}$), and the vulnerable functions of the root \vpsp ($\mathcal{VF}$).
The worklist is initialized with the root {\vpsp}, and the algorithm iteratively processes items in it.

At the beginning of each pass (step \circled{0}), one item ({\vpsp}) is fetched from the worklist.
At step \circled{1}, the algorithm generates the TV list ($new\_tvs$) of current item that are affected by the vulnerability.
Specifically, if it is the first pass, the list is initialized with $\mathcal{VV}$.
Otherwise, the current item must have its own upstream {\vpsp}s, the list of which is saved and updated in the previous passes (step \circled{9}).
Consequently, the algorithm queries the \textit{Inter-PV Calls} record of each upstream {\vpsp} in this list to get the latest affected versions of the current item into $tvs$.
It then loads the cached old version list ($old\_tvs$) if the current {\vpsp} has been processed in previous passes, and derives the difference between the two lists to get the new affected versions ($new\_tvs$).
The combination of $old\_tvs$ and $tvs$ is cached for future use.
At step \circled{2}, the algorithm generates the TF list ($new\_tfs$) of the current item that are affected by the vulnerability.
The generation process is similar to the generation of $new\_tvs$.
Also, the combination of $old\_tfs$ and $tfs$ is cached for future use.
Then, at step \circled{3}, the algorithm queries the dependency graph to extract the dependency relationships between the current item and its direct downstream dependent {\vpsp}s into $pdeps$.

\begin{algorithm}[t]
  \caption{Worklist-based Propagation Algorithm}
  \label{alg:worklist}
  \KwIn{Vulnerable P $\mathcal{VP}$, vulnerable versions $\mathcal{VV}$, vulnerable functions $\mathcal{VF}$}
  \KwOut{All the cached analysis results in \autoref{fig:vpa-detailed}}

\SetKwData{ROOT}{ROOT}
\SetKwData{W}{worklist}
\ROOT $\gets$ \textbf{true}, \W $\gets$ \{$\mathcal{VP}$\}\; 
 
\While{\W is not empty}{
  \circled{0} $item \gets$ \W.pop()\;
  $(tvs, tfs) \gets$ getTVAndTF($item$, $\mathcal{VV}$, $\mathcal{VF}$, \ROOT)\;
  \ROOT $\gets$ \textbf{false}\;
  $(old\_tvs, old\_tfs) \gets$ loadOldTVAndTF($item$)\;  
  \circled{1} $new\_tvs \gets$ diffMergeSave($old\_tvs$, $tvs$)\;
  \circled{2} $new\_tfs \gets$ diffMergeSave($old\_tfs$, $tfs$)\;

  \circled{3} $pdeps \gets$ genPDeps($item$)\;
  // hierarchical pruning: \circled{4} -- \circled{8}\;
  $(s, v3) \gets$ prune($item$, $pdeps$, $new\_tvs$, $new\_tfs$)\;

  \If{$s$ is changed}{
    \circled{9} propagateTVAndTF($item$, $v3$)\;
    \circled{10} \W.extend($v3$)\;
  }
}
\end{algorithm}

With $pdeps$, $new\_tvs$, and $new\_tfs$ available, the hierarchical pruning begins.
At step \circled{4}, the algorithm generates the {\vpspv} dependencies for the current upstream {\vpsp} by querying the \verb|deps.json| records belonging to {\vpsp}s in $pdeps$, and then prunes out irrelevant {\vpspv} dependencies by checking whether the downstream {\vpspv}s rely on upstream {\vpspv}s covered by $new\_tvs$.
After this step, the dependency relationships between upstream {\vpspv}s and the remaining downstream {\vpspv}s are used to generate or update \textit{PV Deps v1}.
At step \circled{5}, the algorithm further prunes dependencies by checking whether the downstream {\vpspv}s import contents from upstream {\vpspv}s in \textit{PV Deps v1}.
The check is restricted to the intrinsic scope for both upstream and downstream {\vpspv}s.
After this step, \textit{PV Deps v2} is generated or updated.
Then, at step \circled{6}, the algorithm identifies new EPs ($new\_eps$) for CG-level pruning.
To achieve this, the algorithm loads the cached old EP list, derives an EP list ($eps$) by constructing CG and conducting backward BFS traversal from $new\_tfs$ to externally accessible functions, and uses the difference between the two lists as $new\_eps$ for each upstream {\vpspv}.
The combination of $old\_eps$ and $eps$ is cached for future use.
Notably, empty $new\_eps$ for all the upstream {\vpspv}s indicates that the dependency state $s$ of the current item has not changed, and the algorithm will not append its downstream {\vpsp}s to the worklist.
With $new\_eps$ available, at step \circled{7}, the algorithm prunes the dependencies by checking whether downstream {\vpspv}s call EPs in $new\_eps$.
After this step, \textit{PV Deps v3} ($v3$ in \autoref{alg:worklist} for brevity) is generated or updated.
At the end of the pruning process, the algorithm generates and caches the \textit{Inter-PV Calls} record for the current item at step \circled{8}.

At step \circled{9}, the algorithm sends the inter-{\vpspv} information (versions and calls) to each downstream {\vpsp} from \textit{PV Deps v3} for future passes.
Finally, at step \circled{10}, the algorithm appends the downstream {\vpsp}s from \textit{PV Deps v3} to the worklist.
The algorithm continues until the worklist becomes empty.

\subsubsection{Example}\label{sec:vpa_example}
We illustrate the algorithm workflow process using CVE-2016-5393~\cite{nistCVE20165393}, 
a high-severity vulnerability in \textit{org.apache.hadoop:hadoop-common}. 
In affected versions of Hadoop, a remote attacker can execute commands with HDFS privileges. 
For brevity, we present only the first round of propagation analysis.
Further propagation results and impact quantification of CVE-2016-5393 are presented in \autoref{sec:case_study}.

At step \circled{0}, \textit{org.apache.hadoop:hadoop-common} is popped as the root {\vpsp}. 
At steps \circled{1}–\circled{2}, its vulnerable versions and functions are set as targets.
At step \circled{3}, the algorithm queries the dependency graph and finds 1,543 downstream {\vpsp}s. 
At step \circled{4}, version-based pruning removes dependents on non-vulnerable {\vpspv}s; 
for example, \textit{com.wgzhao.addax:hbase20xreader:4.0.3} depends on a non-vulnerable version and is pruned, 
leaving 491 {\vpsp}s and 4,258 {\vpspv}s as \textit{PV Deps v1}.
At step \circled{5}, import-based pruning excludes dependents that declare but never import upstream classes; 
for example, although \textit{com.fiftyonezero.eel:eel\-elasticsearch\_2.10:0.11.0} claims that it depends on a vulnerable version of upstream {\vpspv}, it actually does not import any class from this {\vpspv} and is pruned out.
After the pruning, \textit{PV Deps v2} is generated, which contains 373 {\vpsp}s and their 3,321 corresponding {\vpspv}s.
At step \circled{6}–\circled{7}, CG-level pruning removes dependents that import but do not call vulnerable entrypoints; 
for instance, although \textit{com.tencent.angel:angel-ps-graph:2.4.0} depends and imports classes from a vulnerable upstream {\vpspv}, it actually does not call any entrypoint from this {\vpspv} and is pruned out here.
After the pruning, \textit{PV Deps v3} is generated, which contains 228 {\vpsp}s and their 1,685 corresponding {\vpspv}s.
Finally, steps \circled{8}–\circled{10} update the inter-{\vpspv} information and enqueue downstream {\vpsp}s in \textit{PV Deps v3} for the next round.

\subsection{VPSS Calculation}\label{sec:vpss}

After the propagation analysis, we obtain graph-based statistics that reflect a vulnerability’s impact across the ecosystem.
However, these raw data are not readily interpretable or actionable. 
To better profile this impact, we introduce the Vulnerability Propagation Scoring System (VPSS)—a metric that transforms propagation data into a standardized and meaningful impact score for software supply chain vulnerabilities.

VPSS is defined as a propagation-aware measure of vulnerability impact that combines two dimensions: 
the breadth (the affected share of downstream packages) and the depth (the length of dependency chains). 
Higher scores therefore directly reflect wider and deeper propagation in the ecosystem.
The design of VPSS follows three key principles:
(1) Graph awareness: The metric should incorporate both breadth and depth in the dependency graph to capture how widely and deeply a vulnerability propagates.
(2) Interpretability and compatibility: VPSS must be easy to understand, ensuring seamless integration into current vulnerability management workflows and complementing static severity metrics such as CVSS.
(3) Time-awareness: As ecosystems evolve, the metric should adapt to changes such as new dependencies or patches. 
VPSS is thus a dynamic score, supporting longitudinal tracking and timely risk assessment.

To apply these principles, VPSS transforms the results of vulnerability propagation analysis into a normalized impact score within the 0--10 range.
As shown in \autoref{fig:overview}, the score is divided into four tiers—\textit{low} (0--4), \textit{medium} (4--7), \textit{high} (7--9), and \textit{critical} (9--10)—for intuitive risk interpretation.

VPSS captures the breadth and depth of vulnerability propagation through two multiplicative factors: 
\textit{Propagation Breadth Factor} (PBF), which quantifies how widely a vulnerability spreads via direct and transitive downstream dependencies, and \textit{Propagation Depth Factor} (PDF), which measures how deeply it penetrates the dependency graph based on propagation chain length.
These two factors define the raw score:\vspace{-0.5em}

{\small
\begin{align}
\label{eq:vpss_raw}
VPSS_{\text{raw}} = PBF \times PDF
\end{align}
}

The PBF component is computed from four normalized ratios representing the proportion of affected downstream {\vpsp} and {\vpspv} entities, separately for direct and transitive dependencies.
Here, $\text{Total\_P}$ and $\text{Total\_PV}$ denote the total number of {\vpsp}s and {\vpspv}s in the target ecosystem, respectively, which are obtained in the construction process of the dependency graph.\vspace{-0.5em}

{\small
\begin{gather*}
r_{\text{p\_dir}} = \frac{P_{\text{dir}}}{\text{Total\_P}}, \quad
r_{\text{p\_trans}} = \frac{P_{\text{trans}}}{\text{Total\_P}}, \\[1mm]
r_{\text{pv\_dir}} = \frac{PV_{\text{dir}}}{\text{Total\_PV}}, \quad
r_{\text{pv\_trans}} = \frac{PV_{\text{trans}}}{\text{Total\_PV}}
\end{gather*}
}

These values are aggregated using a weighted sum.
Generally, the relationship between these weights should be $w_{\text{p\_dir}} > w_{\text{pv\_dir}} > w_{\text{p\_trans}} > w_{\text{pv\_trans}}$:\vspace{-0.5em}

{\small
\begin{align*}
W &= \begin{pmatrix} w_{\text{p\_dir}} & w_{\text{p\_trans}} & w_{\text{pv\_dir}} & w_{\text{pv\_trans}}\end{pmatrix} \\
X &= \begin{pmatrix} 
r_{\text{p\_dir}} & 
r_{\text{p\_trans}} & 
r_{\text{pv\_dir}} & 
r_{\text{pv\_trans}} 
\end{pmatrix}
\end{align*}
}

To avoid concentration of PBF values in a narrow range, we apply a logarithmic scaling with an amplification factor $\gamma$:\vspace{-0.5em}

{\small
\begin{align}
\label{eq:pbf}
PBF = \ln\left( 1 + \gamma \cdot WX^\top \right)
\end{align}
}

The PDF component is more straightforward, measuring the average and maximum depth of propagation paths on the dependency graph, where $L_{norm}$ is a normalization constant used to adjust the depth metric to a reasonable scale:\vspace{-0.5em}

{\small
\begin{align}
\label{eq:pdf_repeat}
PDF = 1 + \frac{L_{\max} + L_{\text{avg}}}{2 L_{\text{norm}}}
\end{align}
}

The raw VPSS score is then normalized to the final 0–10 range using an exponential saturation function, where $k$ is the saturation parameter controlling the rate at which the raw scores are converted to the ﬁnal scores:\vspace{-0.5em}

{\small
\begin{align}
\label{eq:vpss_repeat}
VPSS = 10 \times \left( 1 - \exp\left( -\frac{VPSS_{\text{raw}}}{k} \right) \right)
\end{align}
}

In total, VPSS introduces seven parameters—$w_{\text{p\_dir}}$, $w_{\text{p\_trans}}$, $w_{\text{pv\_dir}}$, $w_{\text{pv\_trans}}$, $\gamma$, $L_{\text{norm}}$, and $k$—whose values affect the scaling and sensitivity of the score.
Currently, we set these parameters based on domain knowledge and empirical tuning.
We leave automating their setting based on statistical learning from historical vulnerability data as future work.

Lastly, to reflect the evolving nature of software ecosystems, VPSS is explicitly time-aware.
As new software versions are released and patches are applied, the downstream impact of a vulnerability naturally diminishes.
Particularly, when calculating a time-aware VPSS score at $t$, all {\vpsp}s and {\vpspv}s released later than $t$ will be excluded.
Therefore, each VPSS score corresponds to a specific snapshot in time.

\textbf{Rationality Analysis.}
The rationality of VPSS lies in its ability to capture the essential characteristics of vulnerability propagation in software supply chains. 
By combining breadth and depth, the metric reflects that ecosystem risk grows disproportionately when vulnerabilities spread both widely and deeply, whereas either factor alone provides only a partial view. 
The weighting scheme emphasizes direct dependencies and project-level entities, because (1) direct dependencies face greater risks than transitive dependencies due to the different chances of exploitability~\cite{wu2023understanding}, and (2) {\vpsp} dependencies are more stable and reﬂect their impact on the entire ecosystem, while {\vpspv} dependencies may be less stable due to version ﬂuctuations.
Our empirical evaluation ({\autoref{sec:eval_vpss}} and {\autoref{sec:case_study}}) shows that high VPSS values correspond to vulnerabilities with wide and deep propagation (\textit{e.g.,} CVE-2016-5393), while vulnerabilities with limited downstream use receive lower scores.
Moreover, the temporal evolution of VPSS naturally reflects the adoption of patched versions, demonstrating that the metric captures realistic dynamics of vulnerability impact.
These observations demonstrate that VPSS provides a principled and realistic measure of supply chain vulnerability impact.

\section{Implementation}
\label{sec:impl}

We implement an approach prototype for the Java Maven ecosystem in 2.3K lines of Python and 1K lines of Java code.
In this section, we present the implementation details.

\textbf{Dependency Graph Construction.} 
We download the Maven repository index from the Maven Central Repository (MCR)~\cite{apacheCentralIndex}, and parse it with Apache Lucene to extract the {\vpspv} information.
Given the variability of version conflict resolution across projects (\textit{e.g.,} dependency order, exclusions, and overrides), to ensure scalable analysis without requiring build-specific contexts, we choose not to simulate Maven’s full resolution logic, but to use Maven Model Builder to extract the dependency information from the POM files.
We follow prior work~\cite{pashchenko2018vulnerable} to filter out non-deployed dependencies whose \verb|scope| are not \verb|compile| or \verb|runtime|.
Finally, we build the {\vpsp}-level dependency graph with NetworkX, and store it in a Neo4j graph database for efficient querying.

\textbf{Vulnerable Function Identification.}
For the filtering process, we use the GPT-4o-mini API provided by OpenAI to filter out the vulnerability-irrelevant VF candidates.
We also test other mainstream LLMs (\textit{e.g.,} Gemini 2.5 Flash and Claude Sonnet 4) as well as a locally deployed open-source model (Qwen2.5-Coder-32B-Instruct).
Their results are comparable, indicating that our approach benefits from the general reasoning and semantic understanding capabilities of modern LLMs rather than any specific proprietary service.
Since the interface to different models is unified, substituting one for another does not affect the overall workflow.

\textbf{Vulnerability Propagation Analysis.}
We query the dependency graph stored in the Neo4j database to obtain all the downstream dependencies of specific {\vpsp}.
To verify whether a downstream {\vpspv} imports an upstream {\vpspv}, we use Java Dependency Analysis Tool to analyze the JAR files.
To quickly determine whether an upstream method is called by in a downstream {\vpspv}, we utilize Java ASM to parse and search in the bytecode of JAR files.
For CG-level analysis, we use Soot~\cite{vallee2010soot} to analyze JAR files and generate CGs.

\section{Evaluation}
\label{sec:eval}

In this section, we evaluate the effectiveness of our approach by answering the following research questions (RQs):

\textbf{RQ1:} How effective and scalable is our ecosystem-scale vulnerability propagation analysis in identifying potentially affected downstream software projects? (\autoref{sec:eval_vuln_prop})

\textbf{RQ2:} What statistical insights can be drawn from the VPSS scores computed across real-world vulnerabilities? (\autoref{sec:eval_vpss})

All the experiments are conducted on a server equipped with an AMD EPYC 9184X 16-Core Processor and 500 GB of physical memory, running Ubuntu 22.04.5 LTS on the host.

\subsection{Dataset Preparation}

To evaluate the effectiveness of our full-ecosystem vulnerability propagation analysis, we build upon the dataset released by Wu \textit{et al.}~\cite{wu2023understanding}, which contains over 800 vulnerabilities in Maven.
However, running our full analysis on all 800+ vulnerabilities would impose a heavy burden on MCR and significantly increase computational cost.
To balance evaluation thoroughness and practical feasibility, we randomly sample 100 vulnerabilities as our experimental dataset.

\begin{figure*}[t]
    \centering
    \begin{subfigure}[t]{0.3\textwidth}
        \centering
        \includegraphics[width=\textwidth]{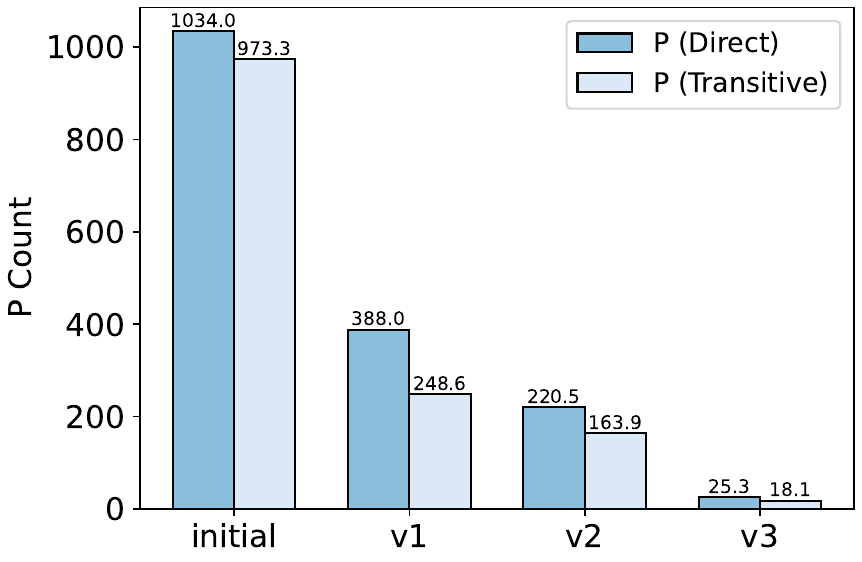}
        \caption{Average P (Direct vs Transitive)}
        \label{fig:p}
    \end{subfigure}
    \hfill
    \begin{subfigure}[t]{0.3\textwidth}
        \centering
        \includegraphics[width=\textwidth]{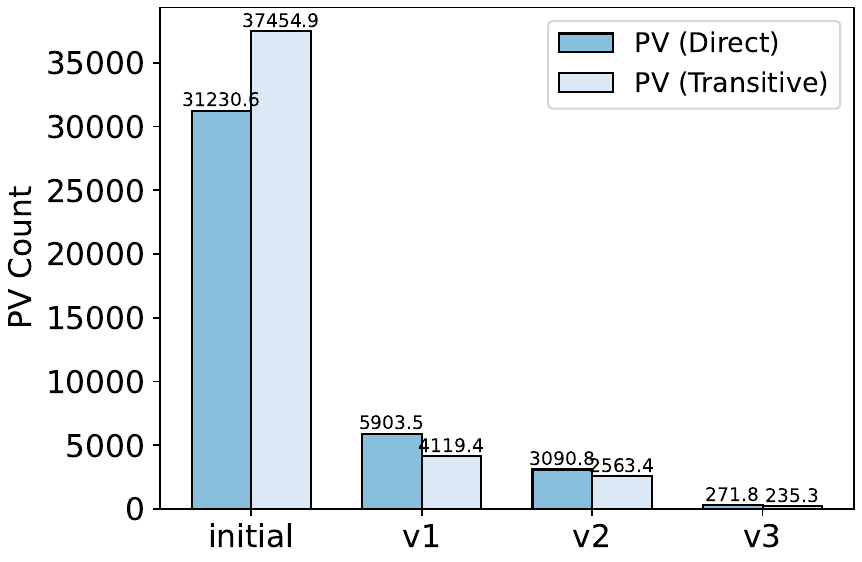}
        \caption{Average PV (Direct vs Transitive)}
        \label{fig:pv}
    \end{subfigure}
    \hfill
    \begin{subfigure}[t]{0.3\textwidth}
        \centering
        \includegraphics[width=\textwidth]{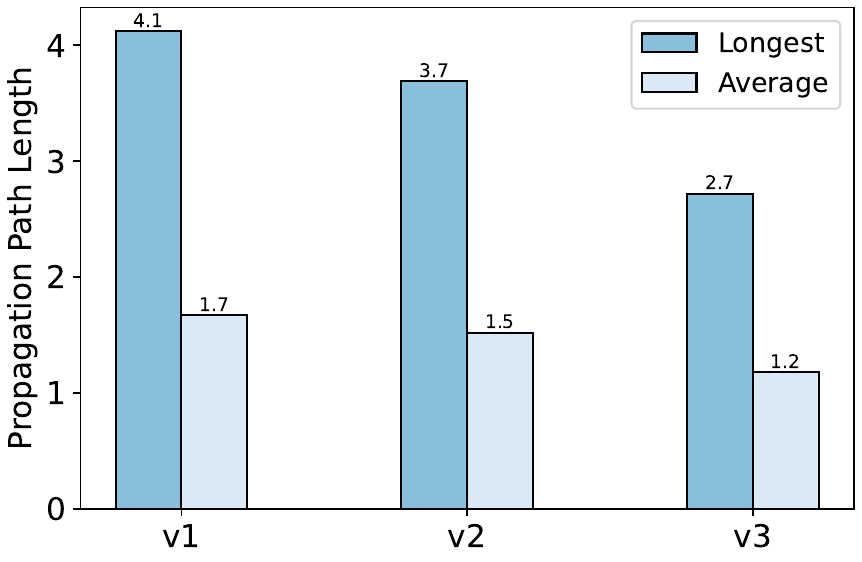}
        \caption{Longest and Average Path Length}
        \label{fig:path}
    \end{subfigure}

    \caption{Average propagation statistics across pruning stages. The v1, v2, and v3 are associated with the pruning results \textit{PV Deps v1}, \textit{PV Deps v2}, and \textit{PV Deps v3} in \autoref{sec:vuln_propagate}, respectively.}
    \label{fig:grouped-avg-metrics}
\end{figure*}

\begin{figure}[t]
  \centering
  \includegraphics[width=\linewidth]{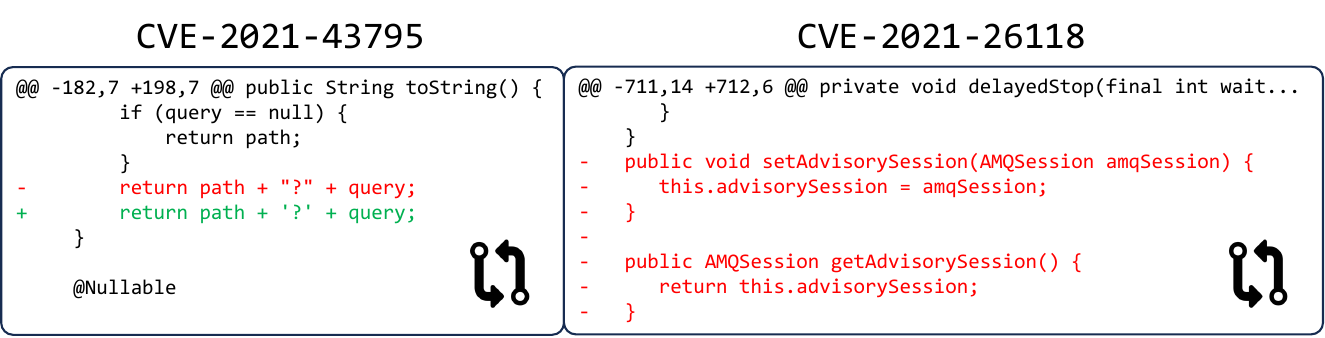}
  \caption{Excerpts of Patches for CVE-2021-\{43795,26118\}}
  \label{fig:fp-vfs}
\end{figure}

With this dataset, we make the following enhancements.
First, in Wu \textit{et al.}'s dataset, only one vulnerable version was selected for each CVE, which can not fully capture the affected version range.
To address this limitation, we enhance the dataset by incorporating the complete list of vulnerable versions for each CVE from the National Vulnerability Database (NVD)~\cite{nvd}.
Second, we identify inaccurate VF annotations in the original dataset with our method in \autoref{sec:vf_identify}, and remove two incorrect VFs.
\autoref{fig:fp-vfs} shows excerpts from the corresponding patches~\cite{patch2021-43795,patch2021-26118}.
In CVE-2021-43795~\cite{nistCVE202143795}, the only change is replacing double quotes with single quotes around the character `?', without altering statement semantics.
Thus, the \texttt{toString()} method should not be considered a VF.
In CVE-2021-26118~\cite{nistCVE202126118}, the removed \textit{Setter} and \textit{Getter} methods for \texttt{AMQSession} only access \texttt{this.advisorySession} and contain no vulnerability-related logic.
In summary, each entry in our refined dataset contains the groupId and artifactId of the vulnerable project, an augmented list of vulnerable versions, and a set of vulnerable functions.

\subsection{Vulnerability Propagation Analysis}
\label{sec:eval_vuln_prop}

For each vulnerability in the dataset, we run our prototype to assess its impact in the Maven ecosystem.
We use the snapshot of MCR index on December 26, 2024, to construct the dependency graph.
There are around 660K {\vpsp}s in the dependency graph, much fewer than the 15M {\vpspv}s in MCR.
During the analyzing process, we limit the request frequency when downloading JAR files to avoid excessive load on MCR.

Excluding the time spent on downloading the JAR packages, the per-vulnerability analysis time ranges from 1.2 seconds to 54 hours, with an average of 5.2 hours and a median of 1.5 hours.
Considering the ecosystem-wide scope of our analysis, this overhead is acceptable and demonstrates the practicality of our approach for large-scale vulnerability impact assessment. 
Moreover, in our measurement each vulnerability was analyzed independently from scratch, whereas in practice, intermediate artifacts such as CGs or intrinsic scope information of {\vpspv}s can be reused across vulnerabilities, which will make future analyses even more efficient.

To further evaluate the effectiveness of our hierarchical worklist-based propagation algorithm, we collect data on the number of {\vpsp}s and {\vpspv}s involved in vulnerability propagation at different stages of the algorithm, as well as the length of the longest propagation paths and the average length of propagation paths on the dependency graph.
\autoref{fig:grouped-avg-metrics} presents the average values of these data at different stages.
For direct and transitive {\vpsp} (\autoref{fig:p}) and {\vpspv} dependencies (\autoref{fig:pv}), we obtain the average values before the whole pruning process and after each pruning step.
For dependency paths (\autoref{fig:path}), we find that getting the average and longest path length before pruning would take too long to compute, so we decide to only look at the results after each pruning step.

We obtain the following findings from the analysis results:
First, the hierarchical pruning mechanism is quite effective, as 97.8\% {\vpsp}s and 99.2\% {\vpspv}s on average are pruned out during the vulnerability propagation analysis.
Also, the length of the longest path and average length decrease at least by 34.1\% and 29.4\%, respectively.
In addition, all the statistics in \autoref{fig:grouped-avg-metrics} decrease in stages, confirming that every pruning process makes its own contributions.
Considering the time and space cost of constructing CGs, our approach greatly reduces the number of CGs that need to be built.
Second, performing vulnerability propagation analysis at the {\vpspv} level based only on project-declared dependencies will result in a large number of false positives, as 94.9\% {\vpspv}s are pruned out with import-based pruning and CG-level pruning.
As shown above, pruning dramatically reduces the number of downstream candidates, which directly translates into fewer CG constructions and proportionally lower analysis time. 
In contrast, a fully non-pruning baseline would require constructing CGs for essentially all downstream packages in the Maven ecosystem, which is computationally infeasible.

\subsection{VPSS Statistics}
\label{sec:eval_vpss}

After the propagation analysis is completed for the dataset, we collect results for VPSS calculation.
As mentioned in ~\autoref{sec:vpss}, we determine the parameter values based on preliminary experiments and expert knowledge, aiming to balance the influence of different components and ensure that VPSS scores meaningfully reflect the propagation impact—higher scores correspond to wider and deeper impact in supply chains.
Particularly, we empirically set the parameters for VPSS computation as follows:
\( w_{\text{p\_dir}} = 5 \), \( w_{\text{p\_trans}} = 2.5 \), \( w_{\text{pv\_dir}} = 3 \), and \( w_{\text{pv\_trans}} = 1.5 \), \( \gamma = 500 \), \( L_{\text{norm}} = 10 \), and \( k = 0.5 \).

\begin{figure*}[t]
  \centering
  \includegraphics[width=0.95\linewidth]{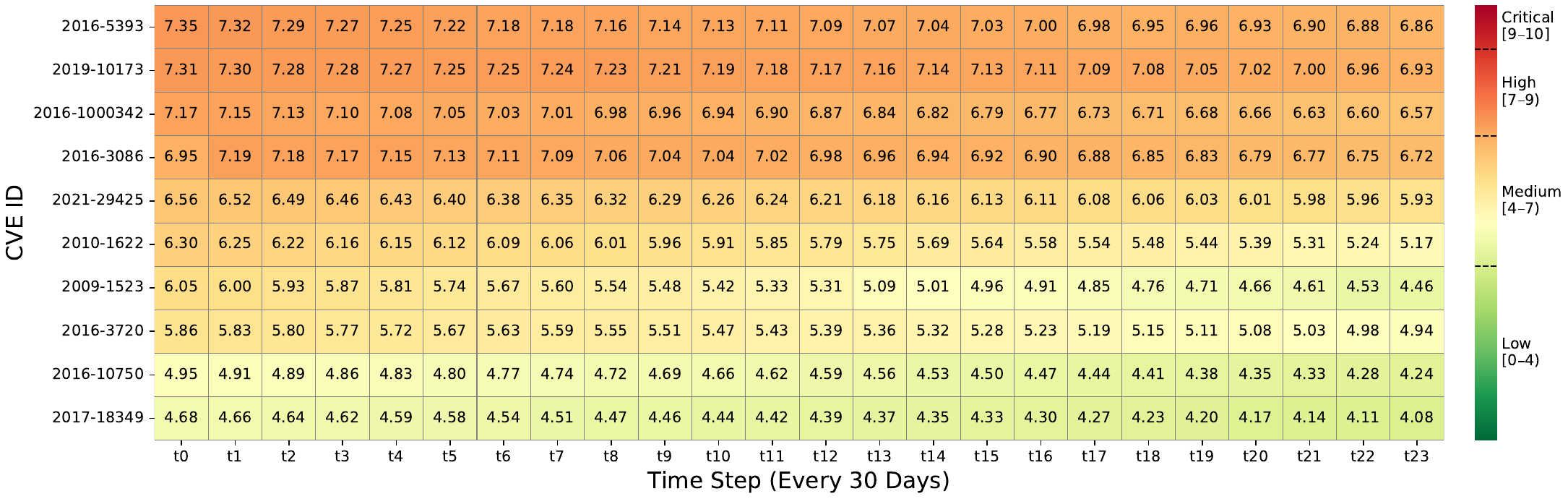}
  \caption{VPSS Time Series (Every 30 Days) for Top-10 CVEs (t0 to t23)}
  \label{fig:top10-vpss-heatmap}
\end{figure*}

\begin{figure}[t]
  \centering
  \includegraphics[width=0.95\linewidth]{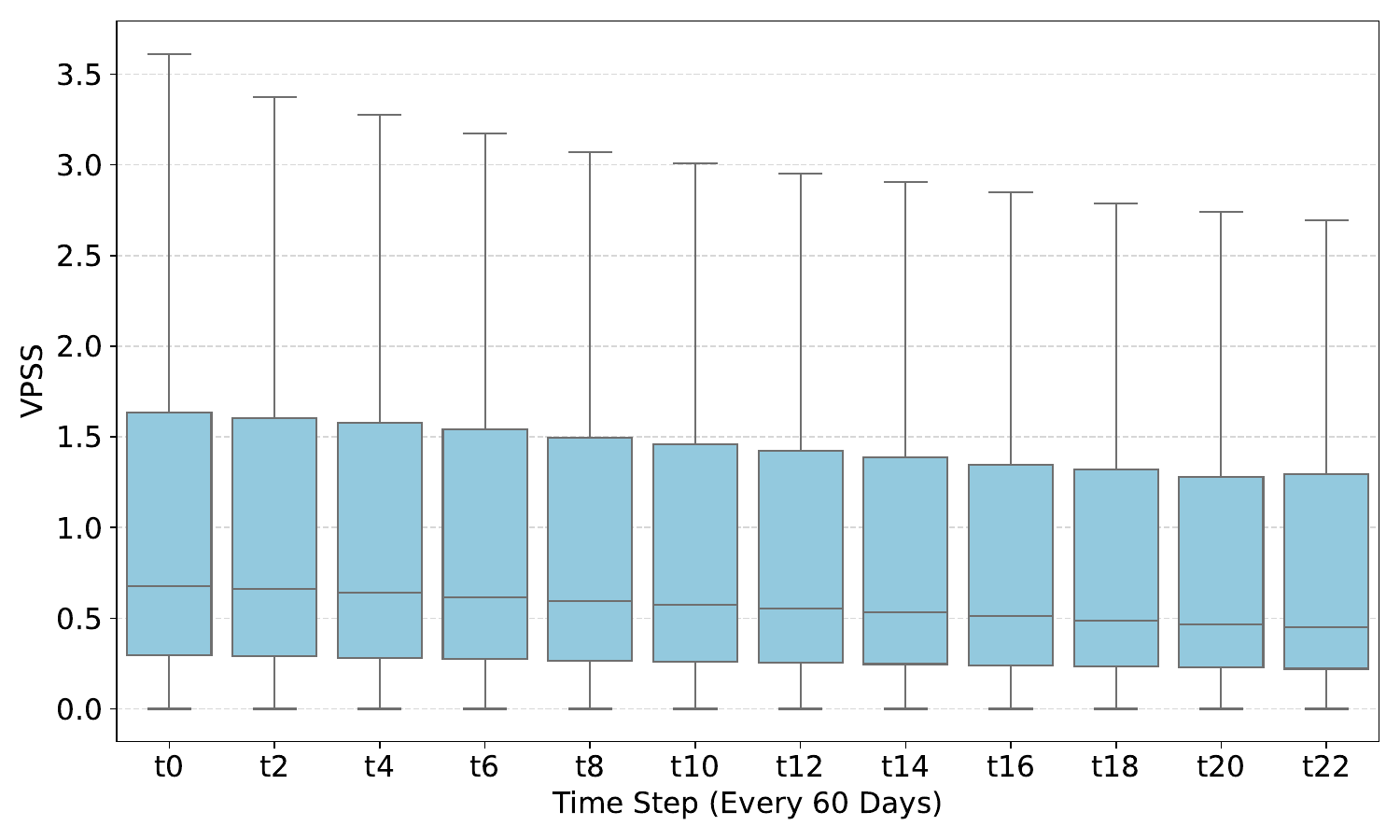}
  \caption{VPSS Distribution (Every 60 Days) Across 100 CVEs}
  \label{fig:vpss-boxplot-60days}
\end{figure}

With such settings, for each vulnerability, we sample 24 time points at 30-day intervals starting from its disclosure date in NVD (denoted as $t_0$).
We compute the VPSS score at each point ($t_0$ to $t_{23}$) to capture the evolution of the vulnerability's impact on software supply chains over approximately 24 months.
To evaluate whether VPSS effectively captures the temporal and distributional characteristics of vulnerability propagation, we visualize its evolution and overall trends.
Specifically, we present \autoref{fig:top10-vpss-heatmap} to highlight VPSS trajectories of the top-10 most impactful CVEs, and \autoref{fig:vpss-boxplot-60days} to show the distribution of VPSS scores across all 100 CVEs over time.

As shown in \autoref{fig:top10-vpss-heatmap}, at $t_0$, the top-10 VPSS scores scale from 7.35 to 4.68, reflecting risk levels from \textit{high} to \textit{medium}.
Nevertheless, the VPSS scores generally decline over time, indicating a decreasing impact on software supply chains.
This trend can be attributed to two main factors: (1) as patched versions are released, an increasing number of downstream projects migrate to vulnerability-free versions; and (2) over time, more projects emerge in the Maven ecosystem.
Additionally, we observe an unusual increase in the VPSS score of CVE-2016-3086~\cite{nistCVE20163086} from $t_0$ to $t_1$ (from 6.95 to 7.19).
This anomaly may be explained by delayed dependency updates in downstream projects~\cite{wu2023understanding}.
Lastly, under the current parameter settings, none of the vulnerabilities in our dataset reach the \textit{critical} VPSS risk level, leaving room for potentially higher-impact vulnerabilities beyond the scope of the current dataset.

Similar to the top-10 CVEs, in \autoref{fig:vpss-boxplot-60days}, the distribution of VPSS scores across all 100 CVEs shows a gradual decline in both median and interquartile range over time.
The boxplots reveal that the majority in the dataset maintain relatively low VPSS scores.
Over time, the overall dispersion narrows slightly, suggesting that the propagation effects of most vulnerabilities tend to stabilize or diminish within two years.
This aligns with the expected lifecycle of patch adoption and ecosystem decoupling from vulnerable packages.

\subsection{Case Study: CVE-2016-5393}
\label{sec:case_study}

This example (CVE-2016-5393~\cite{nistCVE20165393}) shows how our framework captures vulnerability propagation in software supply chains.
Our analysis shows that it exhibits substantial propagation at the ecosystem level.
At $t_0$, the vulnerability directly affected 228 {\vpsp}s and transitively propagated to 154 others, impacting a total of 618 direct and 321 transitive {\vpspv}s.
The resulting VPSS score reached 7.35, which is the highest among all CVEs in our dataset.

Over 24 months, its VPSS score exhibited a gradual decline to 6.86, indicating a slow mitigation pace, which reflects long-tail dependency retention in real-world ecosystems.
Notably, the longest propagation chain consists of 7 dependency hops, spanning critical components of the Hadoop and Hive data processing stacks, illustrating how a single low-level vulnerability can affect a wide range of downstream components.
The average path length also increased slightly over time, reaching 2.33 by $t_{23}$, indicating deepening propagation.

\section{Discussion}
\label{sec:discussion}

\subsection{Accuracy of Propagation Analysis}\label{sec:discuss_accuracy}

There are three main factors that affect the accuracy of our vulnerability propagation analysis.
First, vulnerable versions from public databases such as NVD may be incomplete or inaccurate~\cite{ruan2024kernjc}, leading to false positives or negatives in propagation results.
This can be mitigated by version identification methods~\cite{bao2022v,wu2024vision,cheng2024llm}.
Second, although we enhance patch-based VF identification, real-world security patches occasionally include unrelated but substantial code changes, complicating accurate VF extraction. 
We leave refining VF identification as future work~\cite{sun2025dispatch}.
Third, to achieve ecosystem-scale analysis, we rely on static techniques, \textit{e.g.,} import analysis and CG construction.
While efficient, static analysis may miss paths involving dynamic features (\textit{e.g.,} class loading or method dispatching), causing false negatives. 
Techniques on reflection resolution~\cite{li2019understanding,song2024efficiently} can be integrated into the \textit{import-based pruning} and \textit{CG-level pruning} stages to improve the accuracy of class dependency analysis and CG construction.

\subsection{Adaptability of the Proposed Approach}\label{sec:discuss_adapt}

While our implementation targets the Java Maven ecosystem, the  approach is broadly applicable:
(1) Some Java projects may exist outside of the Maven ecosystem (\textit{e.g.,} hosted only on GitHub).
While we follow prior large-scale studies with Maven~\cite{wu2023understanding,zhang2023mitigating},
our approach does not rely on Maven-specific assumptions: as long as an index of external Java projects can be constructed with dependency metadata, our framework can seamlessly incorporate them into the analysis.
(2) Our framework is ecosystem-agnostic, which can be instantiated in other ecosystems.
For example, in the Python ecosystem, a dependency graph can be built by parsing \texttt{setup.py}, \texttt{requirements.txt}, or \texttt{pyproject.toml}, import relations can be extracted using tools such as pydeps~\cite{githubGitHubThebjornpydeps}, and CGs can be generated via static analyzers like PyCG~\cite{salis2021pycg}.

\subsection{Parameter Setting of VPSS}\label{sec:discuss_vpss}

For VPSS framework (\autoref{sec:vpss}), we introduce a set of configurable parameters to enhance its flexibility across different analytical contexts.
In evaluation, we instantiate these parameters with fixed values based on domain expertise (\autoref{sec:eval_vpss}). 
While this manual configuration suffices for our work, a promising direction for future work is to explore data-driven approaches for parameter tuning.
For example, one could employ statistical optimization or learn optimal settings from historical vulnerability propagation data, enabling more adaptive and context-aware scoring across diverse software ecosystems.

\subsection{Application of VPSS}

The VPSS framework offers three key applications.
First, after a vulnerability is disclosed, VPSS enables quantification of its impact across software supply chains.
It can be used with CVSS to provide more actionable and early‐warning signals.
Second, VPSS can be integrated into vulnerability management workflows to enhance the prioritization process, ensuring that remediation efforts focus on weaknesses with the greatest propagation risk.
Finally, by translating the complex software interdependencies into a standardized score, VPSS facilitates the quantification of software supply chain risk for cyber‐insurance underwriting, supporting more granular and data-driven policy design and premium calculation.  

\section{Related Work}\label{sec:related}

\subsection{Vulnerability Propagation Analysis}

Existing work on vulnerability propagation spans Java, JavaScript, and Python ecosystems:
In Java, Wu \textit{et al.}~\cite{wu2023understanding} and Mir \textit{et al.}~\cite{mir2023effect} perform CG‐level reachability analyses on global and subset Maven graphs;
Ponta \textit{et al.}~\cite{plate2015impact,ponta2018beyond,pashchenko2018vulnerable,pashchenko2020vuln4real,ponta2020detection} combine static and dynamic methods for application‐level VF detection;
Zhang \textit{et al.}~\cite{zhang2024does} determine whether a given project is threatened by vulnerabilities by establishing and querying a vulnerable API database;
others parse POM files for one‐hop dependency analyses~\cite{cadariu2015tracking,kula2018developers,du2018refining} or build dependency graphs for direct and transitive analyses~\cite{hu2019open,wang2020empirical,zhang2023mitigating,ma2024vulnet,shen2025understanding}.
In JavaScript, empirical studies trace client‐side library usage and vulnerability inclusions~\cite{lauinger2018thou}, npm direct dependencies~\cite{decan2018impact}, and ecosystem‐wide propagation via dependency graphs~\cite{zimmermann2019small,wang2023plumber,liu2022demystifying}.
In Python, Ma \textit{et al.}~\cite{ma2020impact} propose a two‐stage impact estimation for scientific projects.

\subsection{Library Vulnerability Exploitation}\label{sec:related_lib_exp}

Beyond propagation analysis, researchers propose methods to generate exploits for library vulnerabilities:
Iannone \textit{et al.}~\cite{iannone2021toward} employ genetic algorithms to evolve test cases that start from clients and reach vulnerable sites in libraries.
Kang \textit{et al.}~\cite{kang2022test} propose to execute vulnerability-revealing tests from libraries to capture triggering states, and guides evolutionary test generation in clients to reproduce these state.
Zhou \textit{et al.}~\cite{zhou2024magneto} combine LLM-guided seed generation with directed fuzzing along call chains to incrementally exploit library vulnerabilities.
Chen \textit{et al.}~\cite{chen2024exploiting} use exploits to guide test generation by migrating crafted parameters into project tests.

\subsection{Vulnerability Assessment}

Researchers have proposed several approaches to improving existing assessments and conducting novel assessments.
Among them, some works aim to automatically predict the CVSS scores~\cite{han2017learning,elbaz2020fighting,le2021deepcva,le2022use,aghaei2023automated,pan2024towards};
some propose to automate the Common Weakness Enumeration classification task~\cite{pan2023fine,fu2023vulexplainer,wen2024livable,luo2024predicting,ji2024applying}.
Additionally, to better understand and profile vulnerabilities, researchers have begun to study more characteristics of them~\cite{ruan2024vulzoo}.
One active area is exploitation prediction~\cite{bozorgi2010beyond,sabottke2015vulnerability,tavabi2018darkembed,chen2019using,jacobs2020improving,jacobs2021exploit,suciu2022expected,jacobs2023enhancing}, which adopts data-driven techniques to estimate the likelihood that a vulnerability will be exploited in the wild.

\section{Conclusion}
\label{sec:conclusion}

This paper fills two key gaps in software supply chain security: the lack of accurate whole-ecosystem vulnerability propagation analysis and the absence of quantitative indicators for assessing vulnerability propagation impact.
We propose a novel framework that combines a hierarchical worklist-based algorithm with multi-level pruning to enable scalable, call-graph-level propagation analysis across direct and transitive dependencies.
We introduce the \textit{Vulnerability Propagation Scoring System} (VPSS), a graph-based metric capturing both propagation breadth and depth over time.
Evaluations on Java Maven ecosystem and real-world CVEs demonstrate the effectiveness of our approach in assessing supply chain risk.

\section*{Acknowledgment}

We thank Zhenxi Li, Shaofei Li, and Yuancheng Jiang for their assistance and the anonymous reviewers for their valuable comments.
This research is supported by the National Research Foundation, Singapore, through the National Cybersecurity R\&D Lab at the National University of Singapore under its National Cybersecurity R\&D Programme (Award No. NCR25-NCL P3-0001). Any opinions, findings and conclusions or recommendations expressed in this material are those of the author(s) and do not reflect the views of National Research Foundation, Singapore, and National Cybersecurity R\&D Lab at the National University of Singapore.

\clearpage
\bibliographystyle{IEEEtran}
\bibliography{references}

\end{document}